\documentclass[preprint,aps,amsmath,byrevtex,prd,titlepage,
nofootinbib]{revtex4-1}

\usepackage{dcolumn}
\usepackage{amsmath}
\usepackage{amssymb}
\usepackage{bm}
\usepackage{hyperref}
\usepackage{graphicx}
\usepackage{slashed}

\newcommand{\beq}{\begin{equation}}
\newcommand{\eeq}{\end{equation}}
\newcommand{\eq}[1]{Eq.~(\ref{#1})}

\begin{document}

\title {Hard Nonlogarithmic Corrections of Order $m\alpha^7$ to Hyperfine Splitting in Positronium}
\author {Michael I. Eides}
%\altaffiliation[Also at ]{the Petersburg Nuclear Physics Institute,
%Gatchina, St.Petersburg 188300, Russia}
\email[Email address: ]{eides@pa.uky.edu, eides@thd.pnpi.spb.ru}
\affiliation{Department of Physics and Astronomy,
University of Kentucky, Lexington, KY 40506, USA\\
and Petersburg Nuclear Physics Institute,
Gatchina, St.Petersburg 188300, Russia}
\author{Valery A. Shelyuto}
\email[Email address: ]{shelyuto@vniim.ru}
\affiliation{D. I.  Mendeleyev Institute for Metrology,
St.Petersburg 190005, Russia}
%\date{}

\begin{abstract}
We consider hard three-loop nonlogarithmic corrections of order $m\alpha^7$ to hyperfine splitting in positronium. All these contributions are generated by the graphs with photon and/or electron loop radiative insertions in the two-photon exchange diagrams. We calculate contributions of six gauge invariant sets of diagrams. The total result for all these diagrams is $\Delta E=-1.2917(1)m\alpha^7/\pi^3=-5.672$ kHz.

\end{abstract}

%\pacs{36.10.Dr,12.20Ds,31.30.jf,32.10.Fn}
%\keywords{hyperfine splitting}

\preprint{UK/14}

\maketitle

%\section{Introduction}

Experimental and theoretical research on hyperfine splitting (HFS) in positronium has a long and distinguished history. The experimental research started with the discovery of positronium \cite{deu1951} and the first HFS measurement \cite{deudu1951}, both in 1951.  Comparable in accuracy results were obtained in the eighties \cite{mb73,rehw84,mills84}

\beq \label{expold}
\Delta E_{exp}= 203~388.5~(1.0)~\mbox{MHz},
\eeq

\noindent
and much later in 2013 \cite{inaksyty2013}

\beq \label{expnew}
\Delta E_{exp}= 203~394.2~(1.6)_{stat}~(1.3)_{sys}~\mbox{MHz}.
\eeq

\noindent
An even later but less accurate result in \cite{mys2014} is compatible with this last number. We see that the recent result in \cite{inaksyty2013} is about three standard deviations higher than the earlier results. In this situation new high precision measurements of the positronium HFS are warranted.

Theoretical work started with calculation of the leading contribution of order $m\alpha^4$ to positronium HFS splitting in the end of forties and beginning of fifties \cite{pir47,ber49,ferr51}. The full quantum electrodynamic  theoretical expression for the positronium HFS splitting is a series in the fine structure constant $\alpha$ with the coefficients that are polynomials in $\ln\alpha$.  During the years many corrections to the leading contribution were calculated. Nowadays all terms up to and including $m\alpha^7\ln\alpha$ are already calculated, see the state of the art theoretical expression in \cite{bmp2014}.  Calculation of the nonlogarithmic corrections of order $m\alpha^7$ is the next theoretical goal. First results for these corrections were published recently \cite{bmp2014,af2014}.

The current theoretical uncertainty of the positronium HFS theory can be estimated by comparison with the known results for HFS in muonium. There are two major differences between HFS in muonium and positronium. First, additional annihilation channel arises in the case of positronium, and second, the masses of constituents coincide in the case of positronium. After calculation of the one-photon annihilation contribution of order $m\alpha^7$ in \cite{bmp2014}, the dominant contribution to the theoretical uncertainty of HFS in positronium is generated by the unknown
nonlogarithmic terms of order $m\alpha^7$  that are similar to the terms of order $\alpha(Z\alpha)^2E_F$ in muonium.  These corrections in muonium are generated by the one-loop radiative insertions in the electron line. Structurally they are similar to the  classic Lamb contributions and are represented by the series in $\ln(Z\alpha)$, $[c_1\ln^2(Z\alpha)+c_2\ln(Z\alpha)+c_3]\alpha(Z\alpha)^2E_F$, see, e.g., review in \cite{egs2001,egs2007}. The nonlogarithmic term with the coefficient $c_3$ arises from the ultrasoft momenta of order $m(Z\alpha)^2$ and is about $2\times 10^{-6}E_F$ \cite{pach1996,kn1997} in muonium. Similar ultrasoft terms in positronium arise due to one-loop radiative insertions in each of the fermion lines and also due to an exchange of an ultrasoft photon between the fermion lines. In this situation we accept $2\times 10^{-6}E^{Ps}_F\sim 250$ kHz as a fair estimate of the ultrasoft nonlogarithmic contribution in positronium.  This term is still uncalculated and after the recent calculation of the one-photon annihilation contribution in \cite{bmp2014} it can be used as an estimate of the theoretical uncertainty of the current positronium HFS theory.

\begin{figure}[htb]
\includegraphics
[height=1cm]
{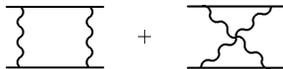}
\caption{\label{twoph}
Diagrams with two-photon exchanges}
\end{figure}

\begin{figure}[htb]
\includegraphics
[height=1.5cm]
{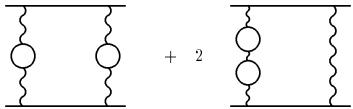}
\caption{\label{twooneloop}
Diagrams with two one-loop polarization insertions}
\end{figure}

\begin{figure}[htb]
\includegraphics
[height=1.5cm]
{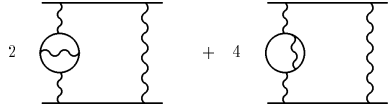}
\caption{\label{onetwoloop}
Diagrams with two-loop polarization insertions}
\end{figure}

\begin{figure}[htb]
\includegraphics
[height=2cm]
{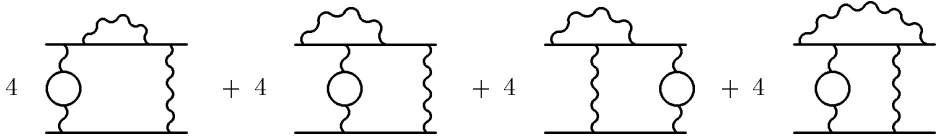}
\caption{\label{onelopradph}
Diagrams with one-loop polarization and radiative photon insertions}
\end{figure}

\begin{figure}[htb]
\includegraphics
[height=2cm]
{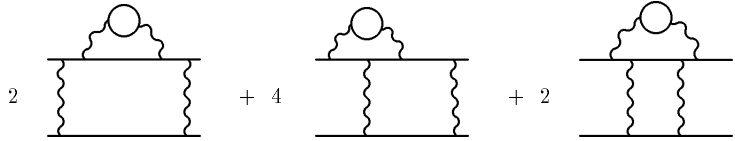}
\caption{\label{polinrad}
Diagrams with one-loop polarization insertions in radiative photons}
\end{figure}

\begin{figure}[htb]
\includegraphics
[height=2cm]
{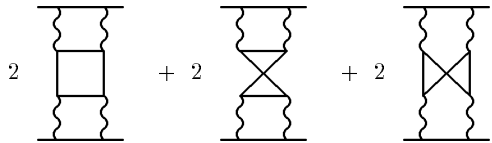}
\caption{\label{lbl}
Diagrams with light-by-light scattering insertions}
\end{figure}

\begin{figure}[htb]
\includegraphics
[height=2cm]
{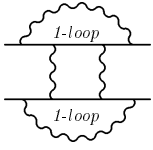}
\caption{\label{combo}
Diagrams with one-loop radiative photon insertions in both fermion lines}
\end{figure}

There are other sources of  nonlogarithmic corrections of order $m\alpha^7$ besides ultrasoft nonlogarithmic contributions. Hard nonlogarithmic corrections of order $m\alpha^7$ are generated by  seven gauge invariant sets of nonannihilation diagrams, six of them are presented in  Figs.~\ref{twooneloop} - \ref{onelopradph} and in Figs.~\ref{polinrad} - \ref{combo}\footnote{We systematically omit diagrams with crossed exchanged photons in all figures.}. One more set of diagrams with two-photon exchanges and two radiative photon insertions in one and the same fermion line also generates hard nonlogarithmic corrections of order $m\alpha^7$. The contributions to HFS in positronium produced by these  diagrams are similar to the contributions of orders $\alpha^2(Z\alpha)E_F$, $\alpha^2(Z\alpha)(m/M)E_F$, and $\alpha(Z^2\alpha)(Z\alpha)(m/M)E_F$ in muonium, see reviews in \cite{egs2001,egs2007,egs2005} and more recent results in \cite{es2009prl,es2009pr,es2010jetp,es2013,es2014}. We report below the results of calculations of gauge invariant contributions to HFS in positronium generated by the diagrams in Figs.~\ref{twooneloop} - \ref{onelopradph} and in Figs.~\ref{polinrad} - \ref{combo}\footnote{The contribution of the diagrams with the light-by-light scattering insertions in Fig.~\ref{lbl} was recently obtained in \cite{af2014}.}. All these diagrams can be obtained by radiative insertions in the diagrams with two-photon exchanges in Fig.~\ref{twoph}. Due to radiative insertions the characteristic integration momenta in all these diagrams are or order of the electron mass $m$, much larger than the characteristic bound state momenta of order $m\alpha$. As a result all these contributions can be calculated in the scattering approximation with the on-shell external electron (positron) lines and the result should be multiplied by the Schr\"odinger-Coulomb wave function squared at the origin. To keep control of the positronium calculations we derived general expressions for contributions of order $\alpha^2(Z\alpha)E_F$ to HFS for a system with constituents with an arbitrary mass ratio. We have checked that these expressions reproduce the contributions to HFS in muonium obtained earlier as series in the small mass ratio \cite{egs2001,egs2007,egs2005,es2009prl,es2009pr,es2010jetp,es2013,es2014}. Then we used the same expressions for calculation of the hard nonlogarithmic corrections of order $m\alpha^7$ in positronium.

We start the calculations with the infrared divergent  contribution to HFS in positronium generated by the two-photon exchange diagrams in Fig.~\ref{twoph} calculated in the scattering approximation. It can be written in the form

\beq  \label{basicintini}
\Delta E=-\frac{\alpha}{\pi}E^{\rm Ps}_F (2m^2)\int \frac{d^4q}{i\pi^2q^4}(2q^2+q_0^2)L_{skel}(q)L_{skel}(-q),
\eeq

\noindent
where $E^{\rm Ps}_F=m\alpha^4/3$ is the leading nonannihilation contribution to HFS in positronium and the factor $L_{skel}$ is defined by the the skeleton electron line factor

\beq
L_{skel}^{\mu\nu}(q)\equiv
-\frac{2q^2}{q^4-4m^2q^2_0}\gamma^{\mu}\hat q \gamma^{\nu}
=2L_{skel}\gamma^{\mu}\hat q \gamma^{\nu}.
\eeq

\noindent
Explicitly, after the Wick rotation we obtain

\beq  \label{basicint}
\Delta E=
\frac{\alpha}{\pi}E^{\rm Ps}_F
\frac{4m^2}{\pi}\int_{0}^{\pi} {d\theta}\sin^2{\theta} \int_{0}^{\infty}
{dq^2}\frac{2+\cos^2{\theta}}{(q^2+4m^2\cos^2{\theta})^2}
\equiv
\frac{\alpha}{\pi}E^{\rm Ps}_F \int_0^\infty dq^2f_p(q).
\eeq

\noindent
Radiative insertions in Figs.~\ref{twooneloop} - \ref{onelopradph} and in Figs.~\ref{polinrad} - \ref{combo} make these diagrams infrared convergent and justify validity of the scattering approximation for their calculation. All corrections calculated below are obtained by some modifications of the basic integrals in \eq{basicintini} and \eq{basicint}.

Consider first the diagrams in Fig.~\ref{twooneloop}  with two one-loop polarization insertions

\beq \label{onelppol}
\frac{\alpha}{\pi}I_1(q)=\frac{\alpha}{\pi}
\int_0^1dv\frac{v^2(1-\frac{v^2}{3})}{4m^2+q^2(1-v^2)}.
\eeq

\noindent
The contribution of the diagrams in Fig.~\ref{twooneloop} is obtained by  insertion of the one-loop photon polarization squared $(\alpha/\pi)^2q^4I_1^2(q)$ in the integrand in \eq{basicint}. After calculations we obtain

\beq \label{onepol}
\Delta E_1=3\frac{\alpha^3}{\pi^3}E^{\rm Ps}_F \int_0^\infty dq^2f_p(q)q^4I_1^2(q)
=\left(\frac{6\pi^2}{35} -
\frac{8}{9}\right)\frac{\alpha^3}{\pi^3}E^{\rm Ps}_F
=0.803~043~294\frac{\alpha^3}{\pi^3}E^{\rm Ps}_F,
\eeq

\noindent
where the factor $3$ before the integral accounts for the multiplicity of the diagrams.

Similarly the contribution of the two-loop vacuum polarization in Fig.~\ref{onetwoloop} can be obtained by the insertion of the two-loop photon polarization $(\alpha^2/\pi^2)q^2I_2(q)$ \cite{ks1955,schw1973} in the integrand in \eq{basicint}

\beq
\Delta E_2=2\frac{\alpha^3}{\pi^3}E^{\rm Ps}_F \int_0^\infty dq^2f_p(q)q^2I_2(q),
\eeq

\noindent
where 2 is the combinatorial factor. The integral representation \cite{schw1973} for $I_2$ is too cumbersome to put it down here. Nevertheless it admits an analytic calculation of the integral above, and we obtain

\beq \label{twpol}
\Delta E_2=\left[-\frac{217}{30}\zeta{(3)} + \frac{28\pi^2}{15}\ln{2}+
\frac{\pi^2}{675} +  \frac{403}{360}\right]\frac{\alpha^3}{\pi^3}E^{\rm Ps}_F
=5.209~219~614\frac{\alpha^3}{\pi^3}E^{\rm Ps}_F.
\eeq

\begin{figure}[htb]
\includegraphics
[height=1cm]
{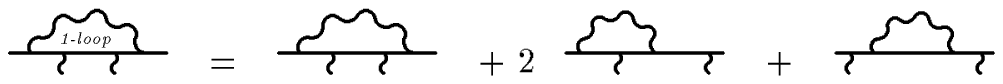}
\caption{\label{ff}
One-loop fermion factor}
\end{figure}

The diagrams in Fig.~\ref{onelopradph} are obtained from the skeleton diagrams in Fig.~\ref{twoph} by one-loop radiative insertions in one of the exchanged photons and one of the fermion lines. Due to the one-loop radiative insertions in the fermion line in Fig.~\ref{onelopradph} effectively the skeleton fermion line in Fig.~\ref{twoph} is replaced by the one-loop fermion factor $L_{\mu\nu}$ in Fig.~\ref{ff}. This corresponds to the substitution

\beq \label{elfact}
L_{skel}^{\mu\nu}(q)
\to L^{\mu\nu}(q)=
2\frac{\alpha}{4\pi}\left\{ \gamma^{\mu}\hat q \gamma^{\nu}
 {\widetilde L}_{\mbox{\tiny I}}(q^2, q^2_0) +q_0
\left[\gamma^{\mu}\gamma^{\nu} -\frac{q^{\mu}\hat q
\gamma^{\nu}+\gamma^{\mu}\hat q q^{\nu} }{q^2}\right]  {\widetilde
L}_{\mbox{\tiny II}}(q^2, q^2_0) \right\},
\eeq

\noindent
in the integral in \eq{basicintini}, where ${\widetilde L}_{\mbox{\tiny I(II)}}(q^2, q^2_0)$  are scalar form factors. We have derived explicit integral representations for these form factors long time ago calculating contributions to HFS in muonium \cite{beks1989,egs1998}.

\noindent
Then we are ready to calculate the contribution of the diagrams in Fig.~\ref{onelopradph} to HFS

\beq \label{polelf}
\begin{split}
\Delta E_3&=
\frac{\alpha^3}{\pi^3}E_F^{Ps} \frac{4m^2}{\pi}\int_0^\infty dq^2q^2I_1(q)\int_0^\pi d\theta\sin^2\theta
L_{skel}\left[(2+\cos^2\theta){\widetilde
L}_{\mbox{\tiny I}} -3\cos^2\theta{\widetilde L}_{\mbox{\tiny
II}}\right]
\\
%\eeq
%\[
&=-1.287~09~(1)~\frac{\alpha^3}{\pi^3}E_F^{Ps},
%\]
\end{split}
\eeq

%\noindent
%where $H_{skel}$ is the same factor as $L_{skel}$ in \eq{elfact} but for the %lower fermion line.

Next we turn to the diagrams in Fig.~\ref{polinrad} with the one-loop polarization insertions in the radiative photon. Effectively these diagrams contain a radiatively corrected electron factor in \eq{elfact}. A photon line with the insertion of a one-loop polarization operator has a natural interpretation as a massive photon propagator, with the mass squared $\lambda^2=4m^2/(1-v^2)$. This propagator should be integrated over $v$ with the weight $(\alpha/\pi)v^2(1-v^2/3)/(1-v^2)$, compare \eq{onelppol}. To obtain the electron factor necessary for calculation of the diagrams in Fig.~\ref{polinrad} we restored the photon mass in the one-loop electron factor in \eq{elfact} and made the substitution above. In this way we obtained an explicit integral representation for this radiatively corrected electron factor. This factor is similar to the radiatively corrected electron factor used in our earlier calculations of the respective contributions to HFS in muonium in \cite{kes1988,es2010jetp}.  All entries in a two-loop fermion factor except the two-loop anomalous magnetic moment decrease at least as $q^2$ when $q^2\to0$. As a result the term with the two-loop  anomalous magnetic moments leads to an infrared divergent contribution  in the integral for the diagrams in Fig.~\ref{polinrad}. This linear infrared divergence indicates existence of a contribution to HFS of the previous order in $\alpha$ that is already accounted for. To get rid of this spurious divergence we subtract the term with the two-loop anomalous magnetic moment from the two-loop electron factor. Then the contribution of the diagrams in Fig.~\ref{polinrad} to HFS  can be written in the form

\beq
\Delta E_4=\frac{\alpha^3}{\pi^3}E_F^{Ps} \frac{2m^2}{\pi}\int_0^\infty dq^2\int_0^\pi d\theta\sin^2\theta
L_{skel}\left[(2+\cos^2\theta){L}^p_{\mbox{\tiny I}} -3\cos^2\theta{L}^p_{\mbox{\tiny
II}}\right],
\eeq

\noindent
where ${L}^p_{\mbox{\tiny I}}$ and ${L}^p_{\mbox{\tiny
II}}$ are the two-loop form factors \cite{kes1988,es2010jetp} similar to the one-loop form factors in \eq{elfact} but with the subtracted anomalous magnetic moment terms. After calculations we obtain

\beq \label{polinrd}
\Delta E_4=-3.154~41~(1)~\frac{\alpha^3}{\pi^3}E_F^{Ps}.
\eeq

Calculation of the light-by-light scattering contribution in Fig.~\ref{lbl} proceeds exactly like calculation of the respective nonlogarithmic radiative-recoil correction to HFS in muonium in \cite{es2014}. The only difference is that we need to let the muon mass to be equal to the electron mass and to restore the terms of higher order in the recoil factor $m/M$ omitted terms in \cite{es2014}. This can be easily achieved by restoring the factor $1/(q^2+4m^2\cos^2\theta)$ instead of $1/q^2$ in the integrand in Eq.(32) of \cite{es2014}. Then the integral for the light-by-light diagrams in Fig.~\ref{combo} acquires the form

\beq
\Delta E_5=
\frac{\alpha^3}{\pi^3} E^{Ps}_F\frac{3m^2}{32\pi}\int_0^\infty dq^2\int_0^\pi d\theta\sin^2\theta\frac{T(q^2,\cos^2\theta)}
{(q^2+4m^2\cos^2\theta)^2}.
\eeq

\noindent
The explicit integral representation for the function  $T(q^2,\cos^2\theta)$ can be found in \cite{es2014}. Calculating this integral we obtain

\beq \label{lbln}
\Delta E_5=-0.706~27~(5)~\frac{\alpha^3}{\pi^3}E^{Ps}_F,
\eeq

\noindent
what coincides with the result obtained recently in \cite{af2014}.

Consider now the diagrams in Fig.~\ref{combo}. These diagrams contain one-loop fermion factors in \eq{elfact} in both fermion lines. As we mentioned above in discussion of the diagrams in Fig.~\ref{polinrad} all entries in a fermion factor except the one-loop anomalous magnetic moment decrease at least as $q^2$ when $q^2\to0$. As a result the product of anomalous magnetic moments leads to an infrared divergent contribution  in the integral for the diagrams in Fig.~\ref{combo}. This linear infrared divergence indicates existence of a contribution to HFS of the previous order in $\alpha$. We need to subtract this lower order contribution. To facilitate this substraction we write the fermion factors (after the Wick rotation) in the form

\beq
{\widetilde
L}_{\mbox{\tiny I}}=L_1+L_A, \qquad
{\widetilde
L}_{\mbox{\tiny II}}=L_{\mbox{\tiny II}}-L_A,
\eeq

\noindent
where we have separated the contribution of the anomalous magnetic moment

\beq
L_A=2L_{skel}=\frac{2}{q^2+4m^2\cos^2\theta}
\eeq

\noindent
to the scalar form factors.

After subtraction of the infrared divergent part we calculate the finite  integral for the contribution of the diagrams in Fig.~\ref{combo} to HFS

\beq \label{combon}
\begin{split}
\Delta E_6
&
=\frac{\alpha^3}{\pi^3}E^{Ps}_F\Biggl\{ \frac{m^2 }{4\pi} \int_0^\infty {dq^2\sin^2\theta d\theta}
\biggl[(2+\cos^2\theta )(L_{\mbox{\tiny I}}L_{\mbox{\tiny I}}+L_A
L_{\mbox{\tiny I}}+L_{\mbox{\tiny I}}L_A)
\\
&
-3\cos^2\theta(L_{\mbox{\tiny I}}L_{\mbox{\tiny II}}+L_A L_{\mbox{\tiny II}}-L_{\mbox{\tiny I}}L_A
+L_{\mbox{\tiny II}}L_{\mbox{\tiny I}}-L_AL_{\mbox{\tiny I}}+L_{\mbox{\tiny II}}L_A)
\\
&
+\cos^2\theta(1+ 2\cos^2\theta)
(L_{\mbox{\tiny II}}L_{\mbox{\tiny II}}-L_AL_{\mbox{\tiny II}}-L_{\mbox{\tiny II}}L_A)\biggr]
+\frac{9}{16}\Biggr\}
\\
&
=-4.739~55~(40)~\frac{\alpha^3}{\pi^3} E^{Ps}_F.
\end{split}
\eeq

Next we collect the results in \eq{onepol}, \eq{twpol}, \eq{polelf}, \eq{polinrd}, \eq{lbln}, and \eq{combon}, and  obtain the total hard contribution to HFS of order $m\alpha^7$ generated by the diagrams in Figs.~\ref{twooneloop} - \ref{onelopradph} and in Figs.~\ref{polinrad} - \ref{combo}

\beq
\Delta E=-3.875~0~(4)\left(\frac{\alpha}{\pi}\right)^3 E^{Ps}_F=-1.291~7~(1)\frac{m\alpha^7}{\pi^3}=-5.672~\mbox{kHz}.
\eeq

Then the total state of the art theoretical prediction for HFS in positronium with account of all known today theoretical contributions \cite{bmp2014,af2014} is

\beq
\Delta E_{theor}=203~391.90~(25)~\mbox{MHz},
\eeq

\noindent
to be compared with the experimental numbers in \eq{expold} and \eq{expnew}. Clearly further reduction of both the experimental and theoretical uncertainties is warranted. On the theoretical side calculation of the still unknown ultrasoft and hard nonlogarithmic contributions  of order $m\alpha^7$ is the next goal. Work on calculation of the remaining hard correction of order $m\alpha^7$ is now in progress, and we hope to report its results in the near future.

\acknowledgments

This work was supported by the NSF grant PHY-1066054. The work of V. S. was also supported in part by the RFBR grant 14-02-00467 and by the DFG grant HA 1457/9-1.

%\end{acknowledgments}

\end{document}